\documentstyle [prl,aps,floats,twocolumn]{revtex}
\input psfig.tex
\begin{document}

\def\ba{\begin{eqnarray}}
\def\ea{\end{eqnarray}}
\def\be{\begin{equation}}
\def\ee{\end{equation}}
\def\({\left(}
\def\){\right)}
\def\[{\left[}
\def\]{\right]}
\def\lagrange {{\cal L}}
\def\del {\nabla}
\def\d {\partial}
\def\Tr{{\rm Tr}}
\def\half{{1\over 2}}
\def\fourth{{1\over 8}}
\def\bibi{\bibitem}
\def\S{{\cal S}}
\def\xx{\mbox{\boldmath $x$}}
\newcommand{\labeq}[1] {\label{eq:#1}}
\newcommand{\eqn}[1] {(\ref{eq:#1})}
\newcommand{\labfig}[1] {\label{fig:#1}}
\newcommand{\fig}[1] {\ref{fig:#1}}
\newcommand\bigdot[1] {\stackrel{\mbox{{\huge .}}}{#1}}
\newcommand\bigddot[1] {\stackrel{\mbox{{\huge ..}}}{#1}}

\twocolumn[\hsize\textwidth\columnwidth\hsize\csname @twocolumnfalse\endcsname
\title{Power Spectra in Global Defect theories of
Cosmic Structure Formation} \author{Ue-Li Pen
\thanks{email:upen\@cfa.harvard.edu}, Uro\v s Seljak
\thanks{email:useljak\@cfa.harvard.edu}, and Neil
Turok\thanks{email:N.G.Turok\@amtp.cam.ac.uk}}\address{${}^*$
Harvard College Observatory, 60 Garden St., Cambridge MA 02138\\
$^\dagger$ Harvard-Smithsonian Center for Astrophysics,
60 Garden St., Cambridge MA 02138\\
% ${}^\dagger$Physics Department, University of Wales Swansea, Singleton Park,
%Swansea SA2 8PP\\ 
${}^\ddagger$DAMTP, Silver St,Cambridge, CB3 9EW, U.K.  }
\date\today 
\maketitle

% Abstract
\begin{abstract}
An efficient technique for computing perturbation power spectra in
field ordering theories of cosmic structure formation is introduced,
enabling computations to be carried out with unprecedented precision.
Large scale simulations are used to measure unequal time correlators of
the source stress energy, taking advantage of scaling during matter and
radiation domination, and causality, to make optimal use of the
available dynamic range. The correlators are then re-expressed in terms
of a sum of eigenvector products, a representation which we argue is
optimal, enabling the computation of the final power spectra to be
performed at high accuracy. Microwave anisotropy and matter
perturbation power spectra for global strings, monopoles, textures and
non-topological textures are presented and compared with recent
observations.
\end{abstract}
\vskip .2in
]

\section{Introduction}
Over the next few years high resolution maps of the cosmic microwave
sky will become available.  These maps will allow competing theories
of cosmic structure formation to be tested with exquisite precision.
The primary quantities of interest for comparing theories to
observations are the power spectra of fluctuations, both for the
microwave sky temperature and the mass density.
The predictions of simple inflationary theories have largely
been worked out, typically to one per cent accuracy, and detailed
comparisons with data are taking place.  The state of the main
rival set of theories, based on symmetry breaking and phase ordering,
has been less rosy.  These theories involve a causal source
comprising the ordering fields and/or defects, which continually
perturbs the universe on ever larger scales. In the inflationary
theories, linear perturbation evolution is all that is needed: but
for the defect theories 
full linear response theory 
is required. The defect sources are in general
`decoherent' \cite{albrecht}, providing additional computational difficulty.
This Letter presents a general solution method for solving
the linear response problem in such models.

\section{Stiff Sources: Measuring Unequal Time Correlators}
Accurate codes have been developed for field evolution in 
different symmetry 
breaking theories. But 
a single simulation cannot simultaneously
resolve all scales of observational relevance.
Thus information 
gathered from simulations must be combined in some way. 
It has been clear for some time that the ideal quantity 
which a) uses all the information present in a simulation, 
b) incorporates the powerful properties of scaling evolution and causality,
and c) preserves all the information needed to compute 
power spectra is the unequal time 
correlator (UETC) of the defect source stress energy tensor $\Theta_{\mu
\nu}$:
\newcommand{\cac}{{\cal C}}
\newcommand{\bk}{{\bf k}}
\be
\langle \Theta_{\mu \nu} (\bk,\tau) \Theta_{\rho \lambda}(-\bk,\tau')
\rangle \equiv {\cal C}_{\mu \nu,\rho \lambda} (k,\tau,\tau')
\labeq{uetc}
\ee
where $\tau$, $\tau'$ denote conformal time, and $k$ comoving wavenumber.
Because the perturbed Einstein-matter equations 
are linear, 
all perturbations are 
determined in terms of the source via appropriate 
Greens functions. Thus in principle
all quadratic estimators of the density perturbations are
determined by (\ref{eq:uetc}).

The unequal time correlators are highly constrained by
causality, scaling  and stress energy conservation. 
Causality means that
the real space correlators of the 
fluctuating part of $ \Theta_{\mu \nu}$ 
must be zero for $r>\tau+\tau'$ \cite{ntcausal}.
Scaling \cite{pst} dictates that in the pure matter or radiation eras
$\cac_{\mu \nu,\rho \lambda}
\propto \phi_0^4 /(\tau \tau')^{1\over 2} c_{\mu \nu,\rho \lambda}
(k\tau,k\tau')$, where $\phi_0$ is the symmetry breaking scale
and $c$ is a scaling function. 
Finally, energy and momentum conservation for the stiff source 
(see e.g. \cite{pst})
provide two linear constraints on the four scalar
components of the source. Any pair determines the other two 
up to possible integration constants. In our work we have found the 
best pair to be
the energy density $\Theta_{00}$ and the anisotropic stress
$\Theta^S$; the energy and momentum conservation 
equations give good behavior for all components  on both
superhorizon and subhorizon scales.  This choice is also favoured
by the fact that
$\Theta_{00}$ and $\Theta^S$, along with the 
vector and tensor components,
$\Theta^V$ and $\Theta^T$
fix all superhorizon 
perturbations in the most direct manner.  However, 
we have also checked that 
other choices give consistent results. 

%In addition, $\Theta_00$ and $\Theta^S$, accompanied by the
%vector and tensor components of the source, $\Theta^V$ and $\Theta^T$, 
%determine in the 
%most direct fashion possible the super-horizon evolution of perturbations.
%Dropping spatial derivatives in the linearised Einstein equations
%the evolution of the trace of the metric perturbation $h$ is completely 
%determined by the vanishing of the pseudoenergy in terms of
%$\Theta_00$, and 
%the traceless part $\tilde {h}_{ij}\equiv {h}_{ij}-{1\over 3} 
%\delta_{ij}h$ is completely determined by the traceless part of
%the spatial stress tensor $\tilde{\Theta}_{ij}$.
%%
%\ba
%h(\tau) &=
% = 2 {\sqrt{1+a} \over a^2} \int^\tau_0 d\tau' {a^3(\tau')
% \over 1+a(\tau')}
%\tau_* 4\pi G \Theta_{00}(\tau') \cr
%\dot{\tilde{h}}_{ij}(\eta)
%&= {1\over a^2(\eta)}\int^\eta_0 d\eta' a^2(\eta') 16 \pi G
%\tilde{\Theta}_{ij}(\eta')\cr
%\ea
%
%where we have used the scale factor 
%First, $\cac$ must be consistent
%with causality: 

% Numerically
%not every combination is stable, and we find that the fastest
%converging variables are $\Theta_{00}$ and $\Theta^S$.

As mentioned, our method uses
scaling and causality to extend the dynamic range
of the numerical simulations.
In the simulations, 
the fields start from uncorrelated, but non-scaling
initial conditions, and evolve towards scaling behaviour.
We evolve the fields to some final time $\tau$
when the system is well into the 
scaling regime.
At this
time we compute the stress energy tensor $\Theta_{\mu
\nu}$, which 
we Fast Fourier Transform and decompose into the 
variables $\Theta_{00}$, $\Theta^S$, $\Theta^V$, and $\Theta^T$.
We then repeat the simulation with identical initial
conditions, computing the same quantities 
at several times $\tau'\leq \tau$. This procedure ensures
that the equal time correlations, which dominate the perturbation 
production, are measured when the system is
closest to scaling. 
We store the isotropic averages
of the correlators as functions of
the wave vector magnitude $k$,
e.g. $\cac_{00,S}(k\tau,\tau'/\tau) = <\Theta_{00}(\bk,\tau)
\Theta^S(-\bk,\tau')>$. Statistical and sampling errors are
small at large $k$ because
many modes contribute, but at small 
$k$ the sampling is sparse and the noise larger. 
In particular the value at $k=0$ represents the mean defect density and
should be
discarded.  In order to increase the resolution for small $k\tau$, we
compute the 
correlators in real space and project them
onto the radial separation $r$.  The fact that
the correlators must vanish for
$r>\tau+\tau'$
allows us to solve for an additive constant,  in effect determining 
the $k=0$ value.
The projected real space correlator is then 
transformed back to Fourier space using the continuum Fourier
transform over the compact support 
$
\cac(k)=4\pi\int_0^{\tau+\tau'} r^2 dr\ \cac(r) {\sin(kr)}/{kr}.
$ %\label{eqn:ift}
%\end{equation}
For small values of $k$ we use the latter equation, while for large values
we use the direct Fourier space computation. For intermediate $k\tau \sim 5$
the two match well, but this procedure allows us to extend our dynamic
range in small $k\tau$ by an order of magnitude.
Note that using causality in this way limits the final time in 
the simulation to $\tau < {L\over 4}$ where $L$ is the box size;
typically we evolve in a $400^3$ box to $\tau= {L\over 8}= 50$
so that there are still many independent horizon volumes in the
box. 

\section{Eigenvector Product Representation}
We have explained how one can devote the full numerical 
power at hand to compute the UETCs alone, without wasting 
computational effort or storage on linear perturbation theory.
Next we shall show how very fast Einstein-Boltzmann solvers 
recently developed \cite{sz} can be used to convert the
UETCs into cosmological power spectra with relatively small
numerical effort and very high precision. 

This is done by
representing the UETCs as 
a sum of eigenvector
products 
\cite{ntcausal}. 
The idea is to 
regard the stress energy correlators 
(\ref{eq:uetc}) as 
`symmetric matrices' with indices $\mu \nu, \tau$ and $ \rho \lambda \tau'$.
In practise, to compute the scalar perturbations we  need the 
auto- and cross-correlators of 
two components (for example  $\Theta_{00}$ and $\Theta^S$), and for
the
vector and tensor perturbations we need the two auto-correlators
of $\Theta^V$ and $\Theta^T$.
Regarded as matrices, the correlators involved 
are symmetric and positive definite
(expectation values of squares) and so the eigenvalues are all
real and positive. Matrix index summation is replaced with
an integral $\int d\tau w(\tau)$ with $w(\tau)$ some chosen weighting
function. 
Often the choice of weighting is naturally dictated by scaling
and dimensional analysis, but in any case the results 
were checked to be independent of it.
For sensible choices of $w(\tau)$ the trace of the correlators
is finite: it follows that there is a 
convergent series of positive eigenvalues which may be used to
index the eigenvector sum.
The correlators can then be expressed as an infinite
sum
\begin{equation}
\cac(k,\tau,\tau')= \sum_i \lambda^i v^i(k,\tau) v^i(k,\tau'),
\label{eqn:evp}
\end{equation}
where
\begin{equation}
\int d \tau' \cac(k,\tau,\tau') v^i(k,\tau') w(\tau') =  \lambda^i
v^i(k,\tau).
\label{eqn:evq}
\end{equation}
The indices labelling the components of the stress tensor 
are implicit. We have found that including the 
largest 15 eigenvectors typically reproduces the
unequal time correlators to better than ten per cent precision: 
the effect of including more than 15 on the final 
perturbation power spectra is negligible at the few per cent level.

We believe the properties of this representation make 
it ideal for the purpose of computing cosmological perturbations.
Namely a) the representation automatically minimises
the `least squares' fit, for a given number of eigenvectors 
b) the eigenvectors individually conserve stress energy
c) the eigenvectors individually vanish as $\tau$ 
goes to zero since the correlators $\cac(k,\tau,\tau')$ vanish for
$\tau \ll\tau'$, making 
specification of the initial conditions
trivial, and d) the contribution of successive eigenvectors 
to the perturbations converges quickly, as a result both of 
the decrease in eigenvalue and the increasingly 
oscillatory nature of the eigenvectors. In particular the
incoherent superhorizon ($k\tau <1$) portion of the correlators
is represented
by an infinite sum of ever more oscillatory eigenfunctions, which
have increasingly little effect on the perturbations.
For the computation we discretise the unequal time correlators
sampled in 256 equal logarithmic spacings in the interval $e^{-6.4}
\le k\tau \le e^{9.6}$.
The individual eigenvectors are  each fed into a
full Boltzmann code \cite{sz}, and the total perturbation 
power spectra 
are then given by the sum of those for individual eigenvectors,
weighted by their eigenvalue.

%In this paper
%\ba
%&\lagrange = {1\over 2} \d_\mu\phi\cdot\d^\mu\phi+  
% \omega^2_\pi\phi^1 + \lambda(\phi\cdot \phi - 1) + {}
%&\nonumber\\
%& {1\over 4}\bigl\{\(\d_\mu\phi\cdot \d_\nu\phi\)
%\(\d^\mu\phi\cdot \d^\nu\phi\)  
%	- \(\d_\mu\phi\cdot\d^\mu\phi\)\(\d_\nu\phi\cdot\d^\nu\phi\)\bigr\}&
%\labeq{Lagrangian}\ea
%\be
%\lagrange = {1\over 2}\bigdot{\phi^a} K^{ab}(\d_i\phi)\bigdot{\phi^b} -
%V(\phi,\d_i\phi)
%+ \lambda(\phi \cdot \phi - 1)\ ,
%\labeq{condensed}
%\ee

\newcommand{\mxi}{{\bf \Xi}}

During the pure matter and radiation epochs, the procedure is 
simplified because the correlators scale, and so are 
represented for all $k$ by a single set of eigenvectors,
functions of $k\tau$. We  incorporate the 
matter-radiation transition by repeating 
the computation of unequal time correlators for 
several different values (typically 20) of $\tau/\tau^*$, where
$\tau^*$ is the  
conformal time at matter-radiation equality $\Omega_r(\tau^*) =
\Omega_m(\tau^*)$.  After the simulations are completed, we
collect the results in a matrix of correlators,
$\mxi(k\tau,k\tau', k\tau^*)$, which are then 
diagonalised to produce a set of eigenvectors for each $k\tau^*$
considered. These eigenvectors smoothly interpolate between
those for the radiation era and those for the matter
era, so the Boltzmann code can use a simple spline interpolation 
between them.
The integration solves the full linearised relativistic Einstein
equations tracking the photons $\delta_r$, baryons $\delta_b$, cold
dark matter $\delta_c$, neutrinos $\delta_\nu$, and their relative
couplings.  The evolution includes the full matter-radiation
transition, finite recombination rate, and other effects.  The
Boltzmann calculations 
are accurate to about 1\%.  

\begin{figure}
\centerline{\psfig{file=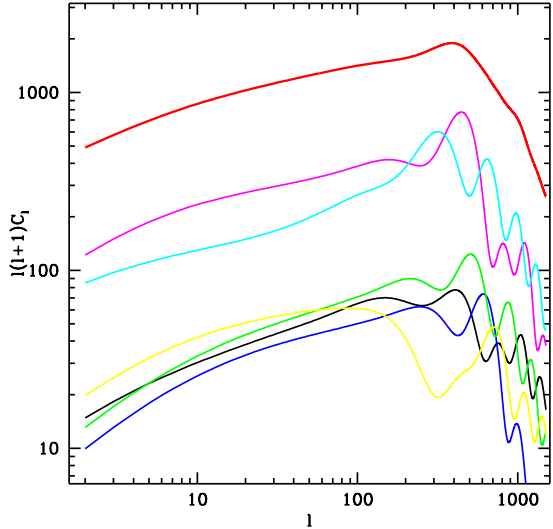,width=3.in}}
\caption{Angular power spectrum of anisotropies generated by the 
scalar component of the source stress energy for global strings. 
The upper curve shows the total spectrum, the lower ones
contributions from individual eigenvectors.
This Figure illustrates decoherence: each eigenvector
individually produces an oscillatory $C_l$ spectrum, but
these oscillations all cancel in the sum.}
\labfig{scalarN2}
\end{figure}

\begin{figure}
\centerline{\psfig{file=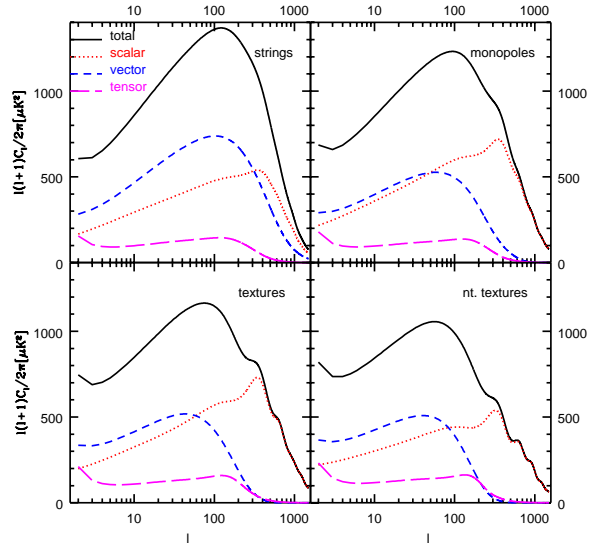,width=3.in}}
\caption{The contributions to the total power from
scalar, vector and tensor components.}
\labfig{comp4}
\end{figure}

\section{Checks}

Many checks have been performed on this procedure, which
we briefly summarise here
\cite{pap2}. 
When the box size
of texture simulations is reduced from $400^3$ to $256^3$, the $C_l$ spectrum
changes by less than 3\%.  To check consistency with 
energy conservation we used a
different pair of scalar variables $\Theta_{00}+\Theta$,
$\Theta+2\Theta^S$, which changed the results by 5\%, 
in $400^3$ boxes, but the latter pair gave much slower
convergence with box size. 
Weighting the diagonalisation by an additional factor of 
$\tau^{1/2}$ adds more weight to subhorizon scales, and affects the
result by up to 10\% at $l=2$, but less than 2\% at larger $l$.  We
have compared the total, as well as scalar, vector and tensor
anisotropies 
to those produced by the direct line-of-sight integration code
\cite{pst}, and they individually agree 
to 10-20\%.  Since the current method includes additional
contributions at the last scattering surface,
explicitly uses 
scaling, employs far larger box sizes, and more accurate integrations, we
conclude the results are consistent with the new results being much
more accurate.  All these checks indicate that these new results are
reliable to better than 10\% \cite{foot}.

\section{Results}
The resulting angular power spectrum of
the temperature anisotropies for the scalar perturbations sourced by
global cosmic strings are shown in Figure \ref{fig:scalarN2}.
Each of the eigenmodes is coherent and shows the expected acoustic
peaks\cite{crit}.  Decoherence is manifested  when we sum over eigenvectors.
The peaks add out of phase, resulting in a smooth angular
power spectrum.  One expects cosmic strings to be the least coherent
of the defects under consideration.  Indeed, the $N=6$ non-topological
texture scalar modes do exhibit residual oscillations even when summed
over all eigenmodes as shown in Figure \ref{fig:comp4}.
The figures show $C_l$ spectra computed
for $h=0.5, \Omega_b=0.05, \Omega=1$.  The dependence on the Hubble
constant $h$ and baryon content
$\Omega_b$ is weak \cite{pap2}.
The most striking feature of all the models under consideration is the
predominance of vector modes.  They dominate up to $l\sim 100$, at
which point they are suppressed by the horizon size on the surface of
last scatter.  It is not hard to see that vector and tensor 
contributions 
should be at least 
comparable to the scalar contribution. By causality
the $k$ space correlator is an integral over a real space
function of compact support, and should be analytic in $k$ at
small $k$.  By
isotropy, we can expand any
correlator at small $k\tau$ as $c_{i
j,k l} (k\tau,k\tau') = A
\delta_{ij}\delta_{kl} + B
(\delta_{ik}\delta_{jl}+\delta_{il}\delta_{jk})
+O(k^2)$.  The constant $A$ contributes only to trace correlators:
the single constant 
$B$ then determines the 
anisotropic scalar, vector and tensor contributions, 
in the ratios 
$c_{S,S} : c_{V,V} : c_{T,T} = 3:2:4$ (these are accurately 
verified in our simulations).  Thus 
vectors and tensors can be expected to 
contribute a significant fraction of
the temperature anisotropies in field ordering theories.
(A
general analysis of the relative importance of vector and tensor 
modes will be described elsewhere
\cite{pap2}.)
The large amplitude of vector modes and the decoherent sum of
eigenmodes leads to a suppression of power at $l\gtrsim
100$ \cite{stebbins}, a very different spectrum to that expected from
adiabatic fluctuations in inflationary models.  
We show a comparison between the predictions of the global field
defect theories and the current generation of CMB experiments in
Figure \ref{fig:cmb}.
All models are normalised to COBE at $l=10$.  They are all 
systematically lower than the current degree-scale experimental points.

\begin{figure}
\centerline{\psfig{file=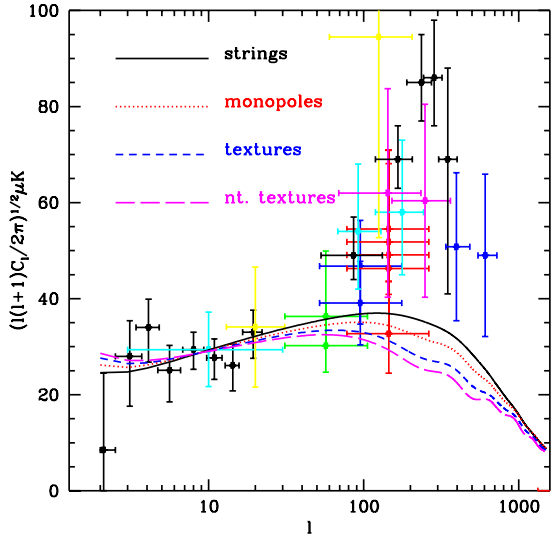,width=3.in}}
\caption{Comparison of defect model predictions to current
experimental data.  All models were COBE normalised at $l=10$.}
\labfig{cmb}
\end{figure}

\begin{figure}
\centerline{\psfig{file=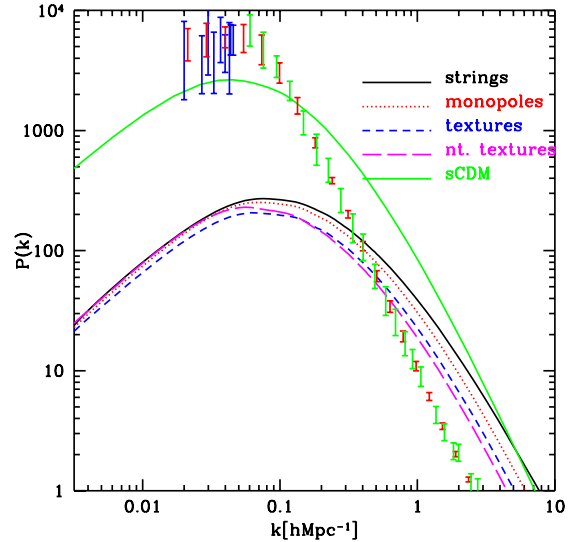,width=3.in}}
\caption{Matter power spectra computed from the Boltzmann code summed
over the eigenmodes.  The upper curve shows the standard cold dark
matter (sCDM) power spectrum.  The defects generally have more power
on small scales than large scales relative to the adiabatic sCDM
model.  The data points show the mass power
spectrum as inferred from the galaxy distribution \protect\cite{peacock}.}
\labfig{matps}
\end{figure}

The same calculations directly yield the
matter power spectrum shown in Figure \ref{fig:matps}.
Normalised to COBE, our tests indicate that the results should be
reliable to a few percent.  
From the power spectra we derive the
normalization $\sigma_8$ of the matter fluctuation in $8h^{-1}$ Mpc spheres.
Global strings, monopoles, texture and $N=6$ non-topological texture
give $\sigma_8=0.26,\ 0.25,\ 0.23,\ 0.21$, respectively, for $h=0.5$, 
and
scaling approximately as $h$. The field normalization for 
textures is $\epsilon=8\pi^2 G\phi_0^2=1.0\times 10^{-4}$, consistent
with our previous calculation\cite{pst} of $\epsilon=1.1\times
10^{-4}$.
These normalizations are a factor of 5 lower than the generic
prediction of $n=1$ inflationary models where $\sigma_8=1.2$ for $h=0.5$.
Cluster abundances suggest values of $\sigma_8 \sim 0.5$ for a flat
universe.

%In conclusion,
%we have developed a procedure to compute perturbations arising from
%stiff, causal sources and applied them to global defects.  
%The only input required is the unequal time correlators of the
%defect stress-energy tensor.  
%
%The
%resolution of the simulations is maximised by only measuring the 
%correlators relative to the end of the run, when noise from the initial
%condition has largely decayed away, and the resolution in units of the
%horizon scale is highest.  We exploit the scaling behaviours of the
%correlators in both matter and radiation epochs and run enough
%simulations to also resolve the matter-radiation transition.  
To summarise, the techniques used here 
enable us to convert unequal time correlators into 
temperature anisotropy and
matter fluctuation power spectra within a few 
hours on a workstation.  For all the defect theories, vectors
contribute approximately half of the total CMB anisotropy on large
scales, leading to a
suppression of acoustic peaks and a low normalization of the matter
power spectrum $\sigma_8 \sim 0.25 h_{50}$.  Current observations of
CMB anisotropies and galaxy clustering do
not favor the models under consideration.

We thank Robert Crittenden, Robert Caldwell and Lloyd Knox for helpful
discussions and John Peacock and Max Tegmark for providing observational
data points. This 
research was 
conducted in cooperation with Silicon Graphics/Cray Research 
utilising the Origin 2000 supercomputer as part of the UK-CCC
facility supported by HEFCE and PPARC (UK).

\end{document}